\documentclass[useAMS]{mn2e}
\usepackage{times}
\usepackage{amssymb}
\usepackage{rotating}
\input{epsf}

\title[Analytic propagating fluctuations]
{An exact analytic treatment of propagating mass accretion rate
  fluctuations in X-ray binaries}
\author[A. Ingram \& M. van der Klis]
{Adam
Ingram$^{1}\thanks{E-mail:a.r.ingram@uva.nl}$ \&
Michiel van der Klis$^{1}$ \\
$^1$Astronomical Institute, “Anton Pannekoek”, University of
Amsterdam, Science Park 904, 1098XH, Amsterdam, The
Netherlands\\
}

\date{Accepted 2013 June 17.  Received 2013 June 15; in original form 2013 June 12}
\pagerange{\pageref{firstpage}--\pageref{lastpage}} \pubyear{2013}

\begin{document}

\topmargin = -0.5cm

\maketitle

\label{firstpage}

\begin{abstract}
Many statistical properties of the aperiodic variability observed in
X-ray radiation from accreting compact objects can be naturally
explained by the propagating fluctuations model. This considers
variations in mass accretion rate to be stirred up throughout the
accretion flow. Variations from the outer regions of the accretion
flow will propagate towards the central object, modulating the
variations from the inner regions and eventually modulating the
radiation, giving rise to the observed linear RMS-flux relation
and also Fourier frequency dependent time lags. Previous treatments of
this model have relied on computationally intensive Monte Carlo
simulations which can only yield an \textit{estimate} of statistical
properties such as the power spectrum. Here, we find \textit{exact}
and \textit{analytic} expressions for the power spectrum and lag
spectrum predicted by the \textit{same} model. We use our
calculation to fit the model of Ingram \& Done (2012) to a power
spectrum of XTE J1550-564. The result we present here will apply to
any treatment of the propagating fluctuations model and thus provides
a very powerful tool for future theoretical modelling.
\end{abstract}

\begin{keywords}
X-rays: binaries -- accretion, accretion discs
\end{keywords}

\section{Introduction} \label{sec:introduction}

Over the last 20 years, mainly thanks to the Rossi X-ray Timing
Explorer (\textit{RXTE}), a detailed phenomenology of the spectral and
timing properties of accreting black hole binaries (BHBs) has emerged
in the literature (see e.g. van der Klis 2006; Done, Gierlinski \&
Kubota 2007; Belloni 2010; Homan et al 2001). A typical transient BHB
in outburst runs through three spectral states classified as the hard,
intermediate and soft states. The hard state spectrum is dominated by
a hard (photon index $\Gamma \approx 1.7$) power law with a
comparatively weak contribution from an optically thick accretion
disc. A softening of the power law ($\Gamma \approx 1.7-2.4$) and an
increased contribution from the disc marks the transition through the
intermediate state, with the disc becoming completely dominant in the
soft state. In addition to direct disc emission, reflection features
are also observed including an iron $K_\alpha$ emission line. The
direct disc emission is well understood as a multi-temperature
black-body (Shakura \& Sunyaev 1973; Novikov \& Thorne 1974) and,
although there is still disagreement about the exact nature of the
accretion geometry, it is generally accepted that the power law
originates from Compton up-scattering of cool `seed' photons (most
likely supplied by the accretion disc, although it is likely that
synchotron radiation from the flow is important in the hard state:
Sobolewska et al 2011; Skipper, McHardy, \& Maccarone 2013) by hot
electrons in some optically thin (optical depth $\tau \approx 1$)
cloud (Thorne \& Price 1975; Sunyaev \& Truemper 1979). The position
and nature of this cloud is the source of much debate but it is often
interpreted as an optically thin accretion flow (hereafter \textit{the
  flow} as opposed to \textit{the disc} surrounding it) formed from
disc evaporation inside some truncation radius, $r_o$ (the
\textit{truncated disc model}: Ichimaru 1977; Esin, McClintock, \&
Narayan 1997; Gilfanov, Churazov, \& Revnivtsev 2000; Done, Gierlinski
\& Kubota 2007; Gilfanov 2010). In the hard state, $r_o$ is large and
thus only a small luminosity of cool disc photons irradiate the flow,
giving rise to a hard spectrum with only a weak direct contribution
from the disc and also weak reflection features. As $r_o$ moves
inward, a greater luminosity of disc photons cools the flow, thus
softening the power law and also increasing the contribution to the
observed spectrum from direct disc emission. The fraction of the power
law emission reflected back into the line of sight from the disc also
naturally increases in this geometry, as is observed (e.g. Gilfanov 2010).

The timing properties of BHBs evolve in a manner tightly correlated
with the spectral evolution. The observed fast ($\sim 100-0.01$s)
variability can be characterised by studying the power spectrum of the
flux time series. Typically the hard state power spectrum consists of
aperiodic broad band noise plus a narrower (Type C) quasi-periodic
oscillation (QPO) with associated harmonics. Phenomenological
modelling using multiple Lorentzian components reveals that all
characteristic frequencies associated with the power
spectrum\footnote{The characteristic frequency is defined as
  $\nu_{peak}^2 = \Delta \nu^2 + \nu_0^2$ where $\Delta \nu$ and
  $\nu_0$ are the width (half width at half maximum) and centroid
  respectively.} \textit{increase} as the source transitions from hard
to soft state, with the QPO becoming increasingly prominent before the
emission becomes stable in the soft state (Belloni, Psaltis \& van der
Klis 2002 and references therein; Churazov, Gilfanov, \& Revnivtsev
2001). In particular, the characteristic frequency of the lowest
frequency Lorentzian in the fit $\nu_b$, often referred to as the low
frequency break, correlates with the QPO frequency (\textit{the
  QPO-break relation}: Wijnands \& van der Klis 1999; also see
Klein-Wolt \& van der Klis 2008). Another fundamental property of the
emission which cannot be probed using the power spectrum is the
\textit{linear} relation between the absolute RMS variability
integrated over any two time scales and the flux averaged over any
longer time scale (\textit{the RMS-flux relation}; Uttley \& McHardy
2001; Heil, Vaughan, \& Uttley 2012). The observation that this
relation seems to hold over all time scales rules out previously
popular shot noise models (Terrell 1972; Weisskopf, Kahn \&
Sutherland 1975), since the RMS and flux from a series of unrelated
flares (or shots) with some shot length drawn from a probability distribution
cannot be linearly related on time scales longer than the shot length
(Uttley, McHardy \& Vaughan 2005).
The variability is also highly coherent across a broad range of energy
bands (Vaughan \& Nowak 1997; Nowak et al 1999) and a cross spectral
analysis reveals Fourier frequency dependent time lags between energy
bands, with hard lagging soft by a greater amount for smaller Fourier
frequencies (Miyamoto \& Kitamoto 1989; Nowak et al 1999). 

Although the physical processes behind the timing properties are very
poorly understood in  comparison to the spectral properties, the broad
band noise is increasingly often attributed to propagating fluctuations in mass
accretion rate (Lyubarskii 1997; Kotov et al 2001; Arevalo
\& Uttley 2006). In this picture, fluctuations stirred up far from the
black hole modulate the mass accretion rate closer to the black
hole. Since this is a multiplicative process, the emitted flux is
naturally predicted to display a linear RMS-flux relation (Arevalo \&
Uttley 2006). If a harder spectrum is emitted from the inner regions
compared with the outer regions, this also naturally gives rise to the
time lags with fluctuations imprinted in the soft band being emitted
in the hard band only after some propagation time (Kotov, Churazov, \&
Gilfanov 2001; Arevalo \& Uttley 2006).

In this model, the power spectral shape of the broad band noise
depends on both the noise generating process \textit{and} the response
of the accretion flow. The magneto-rotational instability (MRI: Hawley 
\& Balbus 1991; Balbus \& Hawley 1998) is most likely the underlying
noise generator, which (very approximately) produces a white noise of
variability everywhere in the flow due to magnetic field lines
interacting with differentially rotating gas. The response of a
Keplerian accretion flow to a white noise of intrinsic fluctuations is
governed by the diffusion equation (Lynden-Bell \& Pringle 1974;
Pringle 1981; Lyubarskii 1997; Churazov, Gilfanov, \& Revnivtsev 2001;
Frank, King \& Raine 2002). Solving this for a $\delta-$function perturbation
(i.e. calculating the Green's function) yields that the power spectrum
of the mass accretion rate far from the radius where the noise
originated is approximately a zero-centred Lorentzian with width
$1/t_{visc}(r)$, where $t_{visc}(r)$ is the local viscous timescale
(Lyubarskii 1997; Pringle 1981)\footnote{Since the Green's function is $\sim
  e^{t/t_{visc}(r)}$ and the Fourier transform of an exponential
is a Lorentzian.}. Since $t_{visc}(r)$ is \textit{longer} for larger
$r$, this implies that variability on different timescales
predominantly originates from different regions of the accretion flow
with higher frequencies coming from closer to the black
hole. Churazov, Gilfanov, \& Revnivtsev (2001), motivated by the
stability of the disc dominated soft state in Cygnus X-1, proposed
that variability is only generated in the flow. In this stable disc /
noisy flow picture, the low frequency break $\nu_b \approx
1/t_{visc}(r_o)$ is naturally predicted to increase as $r_o$ moves
in. Although subsequent observations of disc variability in the hard
state of GX339-4 and SWIFT J1753.5-0127 (Wilkinson \& Uttley
2009; Uttley et al 2011) suggest this picture is overly simplistic, it
forms the starting point for the power spectral model we defined in
Ingram \& Done (2011; 2012; hereafter ID11 and ID12), although we
stress that disc variability must eventually be taken into account.

There is even more uncertainty surrounding the physical origin of the
QPO, with many mechanisms suggested in the literature (Stella \&
Vietri 1998; Markovic \& Lamb 1998; Titarchuk \& Osherovich 1999;
Tagger \& Pellat 1999; Wagoner, Silbergleit \& Ortega-Rodr{\'{\i}}guez
2001; Fragile, Mathews, \& Wilson 2001; Schnittman 2005; Schnittman,
Homan, \& Miller 2006; Cabanac et al 2010). A popular class
of QPO model considers characteristic orbital frequencies at $r_o$. In
General Relativity, frame dragging due to the rotation of a massive
object drags the orbital plane of a test mass around the spin axis of
the massive object, giving rise to precession if the two spin axes are misaligned
(Lense-Thirring precession). Stella \& Vietri (1998) noted that the
observed range of QPO frequencies ($\sim 0.1-10$ Hz) matches the
Lense-Thirring precession frequency of a test mass orbiting at $r_o$
for a range of $r_o$ considered reasonable from spectral fitting
($\sim 60-6~R_g$, where $R_g=GM/c^2$). After Fragile et al (2007)
showed in a general relativistic magnetohydrodynamic (GRMHD)
simulation that an optically thin accretion flow misaligned with the
black hole spin axis can precess as a solid body, Ingram, Done \&
Fragile (2009) suggested that the QPO arises from precession of the
entire inner flow. This turns out to be a very attractive model which
can predict the correct range of QPO frequencies for BHBs (Ingram,
Done \& Fragile 2009) and also atoll sources (low mass accretion rate
neutron star binaries; Ingram \& Done 2010). It also naturally
explains the apparent inclination dependence of QPO strength
(Schnittman, Homan \& Miller 2006) as well as a number of other more
subtle QPO properties (ID11; Sobolewska \& Zycki 2005; Ingram \& Done
2012b; Heil, Vaughan \& Uttley 2011). However in Altamirano et al
(2012), we showed that the $\sim 35-50$ Hz QPO in the 11 Hz pulsar in
the globular cluster Terzan 5 cannot possibly originate from
Lense-Thirring precession. This source displays a QPO-break relation
consistent with other Z-sources, implying that the low frequency QPOs
in Z-sources (horizontal branch oscillations) do not originate from
Lense-Thirring precession. Since Z-sources display a somewhat
different QPO-break relation to atolls and BHBs (following a track a
factor $\sim 2$ higher in QPO frequency and displaying a turn-over at
$\nu_b\sim10$ Hz if data from Terzan 5 X-2 and Sco X-1 are
considered), we cannot rule out the model in general from this
observation.

ID11 and ID12 defined and developed a power spectral model
which combines propagating mass accretion rate fluctuations with
Lense-Thirring precession. The algorithm of Timmer \& Koenig (1995;
hereafter TK95; also see Davies \& Harte 1987) was used to simulate
mass accretion rate fluctuations with a random phase in order to
\textit{estimate} the power spectrum. This led to two major
disadvantages: 1) the power spectrum calculated in this way is inexact
giving rise to errors associated with the model; 2) the simulation is
\textit{very} computationally intensive, meaning it took weeks to find
a local minimum in $\chi^2$ and full error calculations were not
feasible. In this paper, we describe how the same calculation can be
done analytically. We show that this gives the same results as the
simulation and test a slightly modified version of the model against a
data set previously considered in ID12. The new calculation is
extremely fast, meaning we can now fully explore parameter space. We
also discuss how our result here can be combined with more
sophisticated physical assumptions in future work to define more
realistic analytic models of the accretion flow.

\section{A random walk on the complex plane} \label{sec:rwalk}

In the propagating fluctuations model, the mass accretion rate at some
point in the flow is the product of many stochastic time
series. Previous treatments of this model have involved simulating
time series before estimating the power spectrum of their product by
averaging over many realisations. In this section, we show that the
power spectrum of this product can be calculated analytically. We
first define important quantities before considering multiplying two,
followed by an arbitrary number of time series.

\subsection{Definitions}

We define the discrete Fourier transform (DFT) of a time series $a_k$,
evaluated at time $k~dt$ with $k = 1,..,N$, as:
\begin{equation}
A_j = \frac{1}{N} \sum_{k=1}^{N} a_k e^{i2\pi jk/N}
\label{eqn:dft}
\end{equation}
(see e.g. Oppenheim \& Schafer 1975; van der Klis 1989; Press et al
1992). Here, $A_j$ is evaluated at frequency $\nu_j = j~ d\nu = j/(N dt)$
where $j = -N/2+1,..,N/2$. From this it follows that the inverse
transform is:
\begin{equation}
a_k = \sum_{j=-N/2+1}^{N/2} A_j e^{-i2\pi jk/N}.
\label{eqn:inverse}
\end{equation}
Since we are always considering $a_k$ to be some physical signal, it
must be real and thus its DFT is complex conjugate symmetric
($A_{-j}=A^*_j$) and its periodogram, $|A_j|^2$ (again with
$j=-N/2+1,..,N/2$), is symmetric about $j=0$
($|A_{-j}|^2=|A_j|^2)$. Hereafter, we refer to $j=0$ and $j\neq 0$
terms respectively as the DC and AC components (standing for direct
and alternating current). Under these definitions, a time series with
mean $\mu$, variance $\sigma^2$ and duration $T = N dt$, has a DC
component $A_0 = \mu$ and AC components obeying the following form of
Parseval's theorem:
\begin{equation}
\sum_{j=1}^{N/2} |A_j|^2 d\nu - \frac{|A_{N/2}|^2 d\nu}{2} =
\frac{\sigma^2}{2T} .
\label{eqn:parseval}
\end{equation}
For plots in this paper, we re-normalise the periodogram by a factor
$2T$ or $2T/ \mu^2$ such that its integral over all positive
frequencies is approximately $\sigma^2$ or $(\sigma/\mu)^2$. From
equation \ref{eqn:parseval}, we see that this approximation becomes
very good for large $N$.

In our application, $a_k$ is stochastic and so represents a particular
realisation of an underlying process. The periodogram, $|A_j|^2$,
represents the power spectrum of the \textit{realisation}. The average
periodogram, $<|A_j|^2>$, provides an estimate for the power spectrum
of the \textit{process}, $|A(\nu_j)|^2$, which becomes exact when the
averaging is over infinite realisations. Hereafter, we adopt the
convention that \textit{the power spectrum} always refers to the
process and \textit{the periodogram} always refers to the
realisation. The TK95 algorithm generates a realisation of the process
with power spectrum $|A(\nu_j)|^2$. It generates a real time series
$a_k$ as the inverse DFT of the complex conjugate symmetric series
$A_j$, which obeys:
\begin{equation}
<|A_j|^2> = <\Re A_j ^2> + <\Im A_j ^2> = |A(\nu_j)|^2.
\label{eqn:condition}
\end{equation}
For $j = 1,..,N/2-1$, the real and imaginary parts of $A_j$ are random
variables chosen from a Gaussian distribution with zero mean and the
same variance which we can see from equation \ref{eqn:condition} must
be equal to $|A(\nu_j)|^2/2$. Finally, $\Re A_{-j} = \Re A_j$ and
$\Im A_{-j} = -\Im A_j$ to ensure $a_k$ is real. These conditions
ensure that the phase $\phi_j$ is uniformly random on the interval
$-\pi < \phi_j \leq \pi$ for these frequencies. For the Nyquist
frequency ($j=N/2$), $A_j$ is always real. Consequently, the variance of the
zero-mean Gaussian distribution from which $\Re A_j$ is chosen must,
from equation \ref{eqn:condition}, be equal to $|A(\nu_j)|^2$. Even
though the Nyquist component is always real, its phase is still
random, either taking the value $\phi_{N/2}=0$ or $\phi_{N/2}=\pi$,
with an equal chance of each eventuality. We give the time series a
mean, and thus DC component, of $\mu_a$.

Finally, we also define the cross spectrum between two real time series
$a_k$ and $b_k$ as $C(\nu_j) = A(\nu_j)^*B(\nu_j)$ with
$j=-N/2+1,..,N/2$. In direct analogy to our definition of the power
spectrum, we adopt the convention that this is a property of the
process as opposed to a particular realisation and is given by
$C(\nu_j) = <A_j^*B_j>$, with the property $C(\nu_{-j}) =
C(\nu_j)^*$. Note that, although the power and cross spectra are defined
for $j<0$, their symmetry about $j=0$ ensures that the negative
frequency components contain no extra information.

\subsection{Multiplying two time series}
\label{sec:2rings}

\begin{figure}
\centering
\includegraphics[height=5.0cm,width=8.0cm,trim=3.0cm 3.0cm 3.0cm 2.0cm,clip=true]{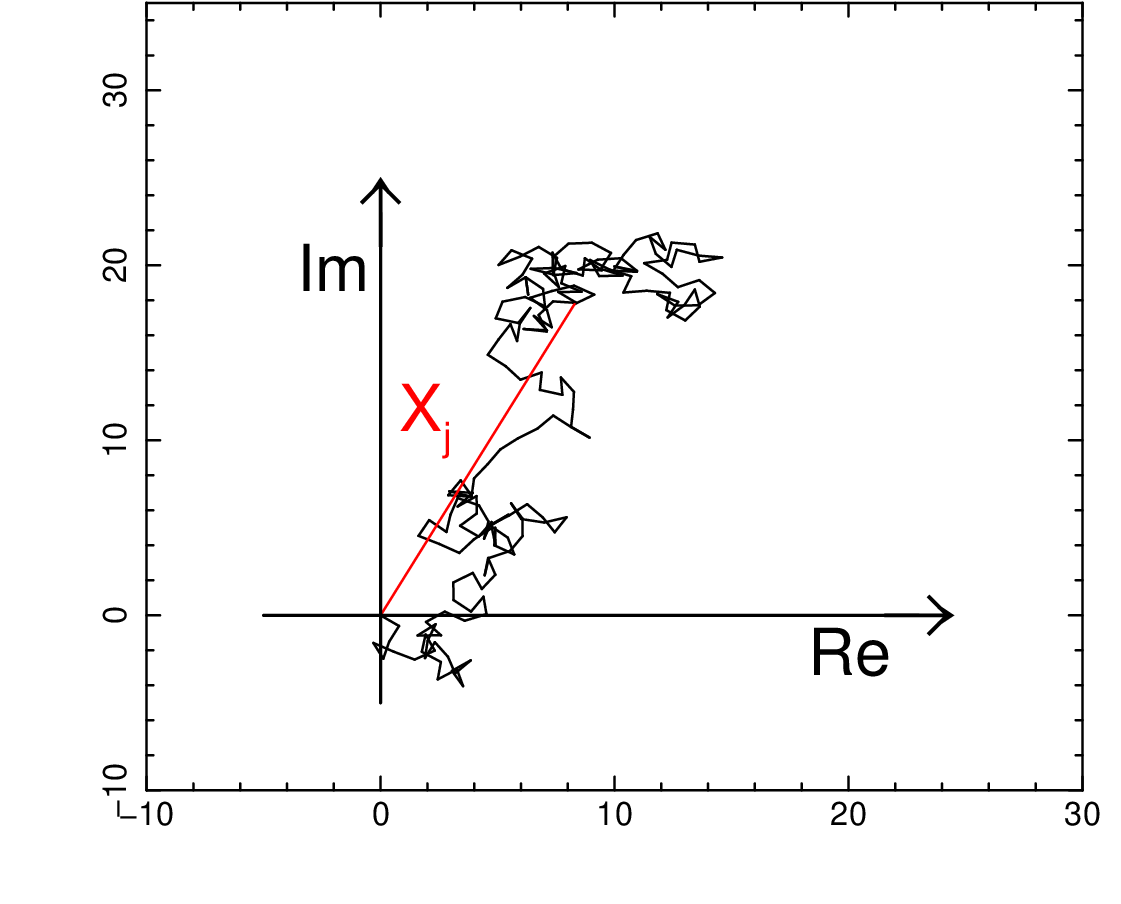}
\caption{Example of a random walk with $N=200$ steps each with length
  $|s_{jk}|=1$ in a random direction. The red line represents the
  complex number $X_j$ which describes the final position of the walk
  after $N$ steps.}
\label{fig:complexplane}
\end{figure}

Let us now take two time series generated via the TK95 method, $a_k$
and $b_k$, and multiply them together to get $x_k = a_k b_k$. The DFT
of $x_k$ is:
\begin{equation}
X_j = \sum_{k=-N/2+1}^{N/2} A_{j-k}B_k .
\label{eqn:cycconv}
\end{equation}
This is the convolution theorem. Since $A_{j-k}$ is periodic on the
interval $j-k=-N/2+1,..,N/2$, we can add or subtract $N$ to the
index of $A$ in order to keep it in the desired interval. We wish to
calculate the power spectrum of $x_k$ without simulating. Since, for
each value of $j$, $s_{jk}=A_{j-k}B_k$ is a series of $N$ random
variables, each with well defined average amplitude and random phase,
equation \ref{eqn:cycconv} represents a random walk on the complex
plane\footnote{Note that the complex conjugate symmetry introduced by
  requiring real time series actually introduces a correlation between
  $s_{jk}$ and $s_{j(-k)}$ terms. However, we show in Appendix
  \ref{sec:evilterms} that the resulting pairs of terms behave exactly
  as if they were uncorrelated.}. $X_j$ is thus the final position on
the complex plane after $N$ steps have been taken. It follows that
$|X_j|$ is the distance covered by the random walk. Figure
\ref{fig:complexplane} shows an example of this with all $|s_{jk}|=1$.

We see from equation \ref{eqn:cycconv} that $|X_j|^2$ is given by a
sum over all $|s_{jk}|^2$ plus many cross terms. If we average
$|X_j|^2$ over infinite realisations, all of these cross terms go to
zero since the $s_{jk}$ terms are uncorrelated with one another and we
arrive at a well known theorem for the length of a Gaussian random
walk (see e.g. Weiss 1994):
\begin{equation}
<|X_j|^2> = \sum_{k=-N/2+1}^{N/2} <|s_{jk}|^2>.
\label{eqn:rwalk}
\end{equation}
Substituting the definition for $s_{jk}$ and using the fact that
$|A_{j-k}B_k|^2=|A_{j-k}|^2|B_k|^2$, we find:
\begin{equation}
<|X_j|^2> = \sum_{k=-N/2+1}^{N/2} <|A_{j-k}|^2> <|B_k|^2> .
\label{eqn:proof1}
\end{equation}
Since the averaging here is over infinite realisations, we can write
this in terms of power spectra:
\begin{equation}
|X(\nu_j)|^2 = |A(\nu_j)|^2 \otimes |B(\nu_j)|^2 ,
\label{eqn:proof}
\end{equation}
where $\otimes$ denotes a convolution. We thus have an expression
to obtain $N/2+1$ values of the analytic function $|X(\nu)|^2$ from
$N/2+1$ values of the analytic functions $|A(\nu)|^2$ and $|B(\nu)|^2$!

\begin{figure}
\centering
\leavevmode  \epsfxsize=8.5cm \epsfbox{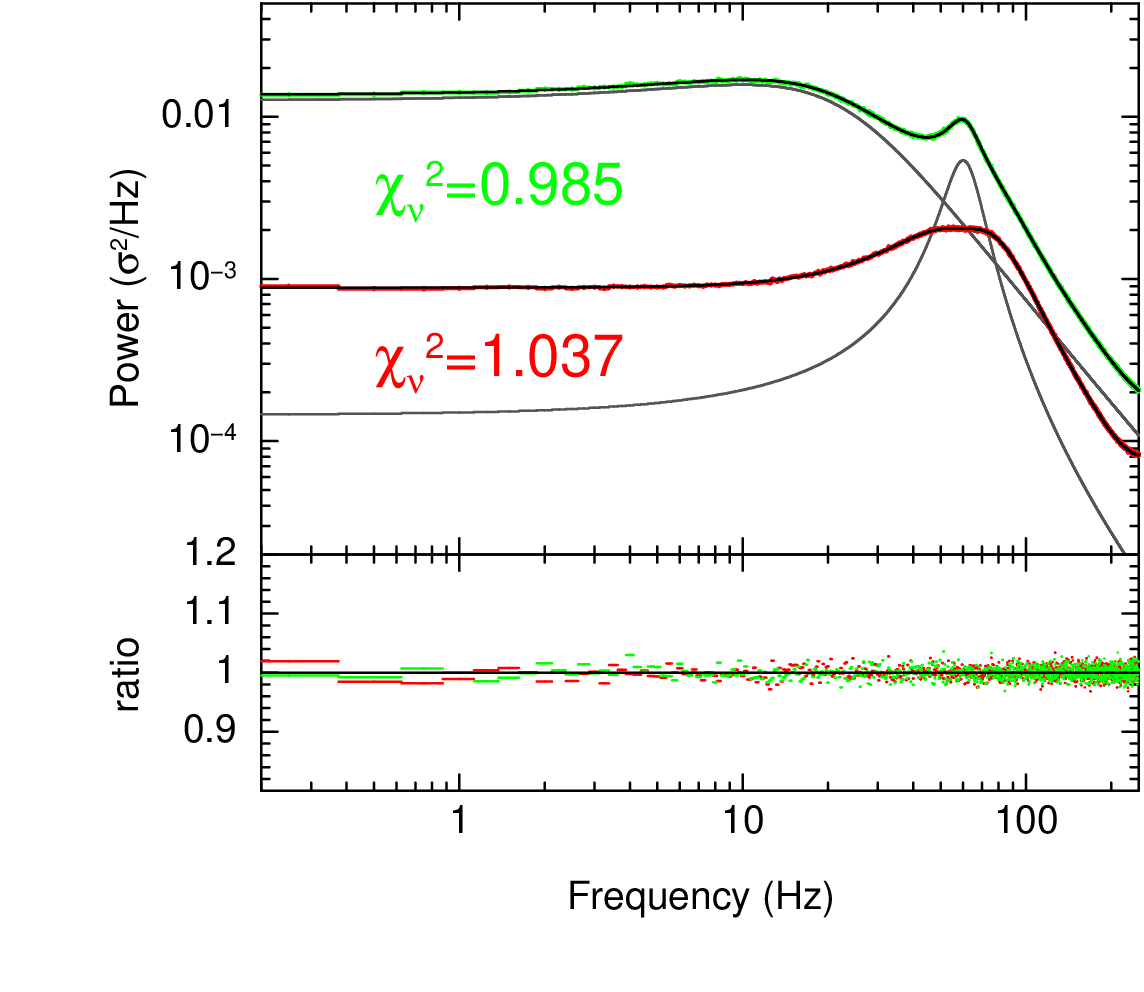}
\vspace{-10mm}
\caption{
Power spectrum of the time series $x_k=a_kb_k$, where $a_k$ and $b_k$
are stochastic time series with average power spectra given by the
grey lines. The red and green points show the power spectrum recovered
from simulating $a_k$ and $b_k$ with a mean of zero and unity
respectively (The green points are above the red points for readers in
black and white). The black lines passing through the simulation results
are analytic calculations. The ratio plots (simulation over
calculation) and $\sim$unity reduced $\chi^2$ values confirm that the
simulation and calculation agree.}
\label{fig:pop}
\end{figure}

As a demonstration, we compare our analytic calculation to simulations. We
take our two input power spectra to be Lorentzians with width $\Delta
\nu =$20 Hz and 10 Hz and centroid $\nu_0=$10 Hz and 60 Hz
respectively. They are normalised to have a standard deviation in the
time domain of $\sigma=$0.8 and 0.4 respectively (see van Straaten et
al 2002) and are shown in grey in Figure \ref{fig:pop}. We use
$N=2^{11}$ and $dt=2^{-9}$ s and simulate 10000 time series using the
TK95 algorithm before estimating the power spectrum by averaging over
all realisations (with an error given by the statistical error on the
mean, yielding 1 sigma error bars). The red and green points in Figure
\ref{fig:pop} show the result of the simulation when the time series
have means of $\mu=$0 and 1 respectively (the $\mu=$1 points are above
the $\mu=$0 points). The two black lines passing
through the simulation points represent the same cases calculated
analytically. The ratio plot (simulation divided by calculation) and
the $\chi^2$ confirm that the two methods give the same result for
both cases. Note, throughout this paper $\chi^2_\nu$ represents
reduced $\chi^2$ (i.e. $\chi^2$ / degrees of freedom).

To understand why changing the input DC components makes such a
difference to the final power spectra in Figure \ref{fig:pop}, we can
re-write equation \ref{eqn:proof} as:
\begin{eqnarray}
|X(\nu_j)|^2 &=& |\tilde{A}(\nu_j)|^2 \otimes |\tilde{B}(\nu_j)|^2 +
\mu_b^2 |\tilde{A}(\nu_j)|^2 \nonumber  \\
 & & + \mu_a^2 |\tilde{B}(\nu_j)|^2
+ \mu_a^2 \mu_b^2 \delta_{j0} ,
\label{eqn:dc}
\end{eqnarray}
where $\delta_{j0}$ is a Kronecker delta and $\tilde{A}_j=A_j$,
$\tilde{B}_j=B_j$ except for $\tilde{A}_0 = \tilde{B}_0 =
0$. This shows that changing $\mu_a$ and $\mu_b$ to unity for the
green line in Figure \ref{fig:pop} has the effect of adding the input
functions to the convolution for $\mu_a = \mu_b = 0$ (i.e. the green
line is the sum of the red line and the two grey lines).

\subsection{Multiplying many time series}
\label{sec:nrings}

The argument of the previous section can be extended to consider
$\mathcal{N}$ time series $(a_n)_k$ being multiplied together to get
$x_k = \prod_{n=1}^{\mathcal{N}} (a_n)_k$. Equation \ref{eqn:proof}
can be used $\mathcal{N}-1$ times to write the power spectrum of $x_k$
as:
\begin{equation}
|X(\nu_j)|^2 = \coprod_{n=1}^{\mathcal{N}} |A_n(\nu_j)|^2,
\label{eqn:coprod}
\end{equation}
where we adopt the co-product symbol to represent a succession of
convolutions.

\begin{figure}
\centering
\leavevmode  \epsfxsize=8.5cm \epsfbox{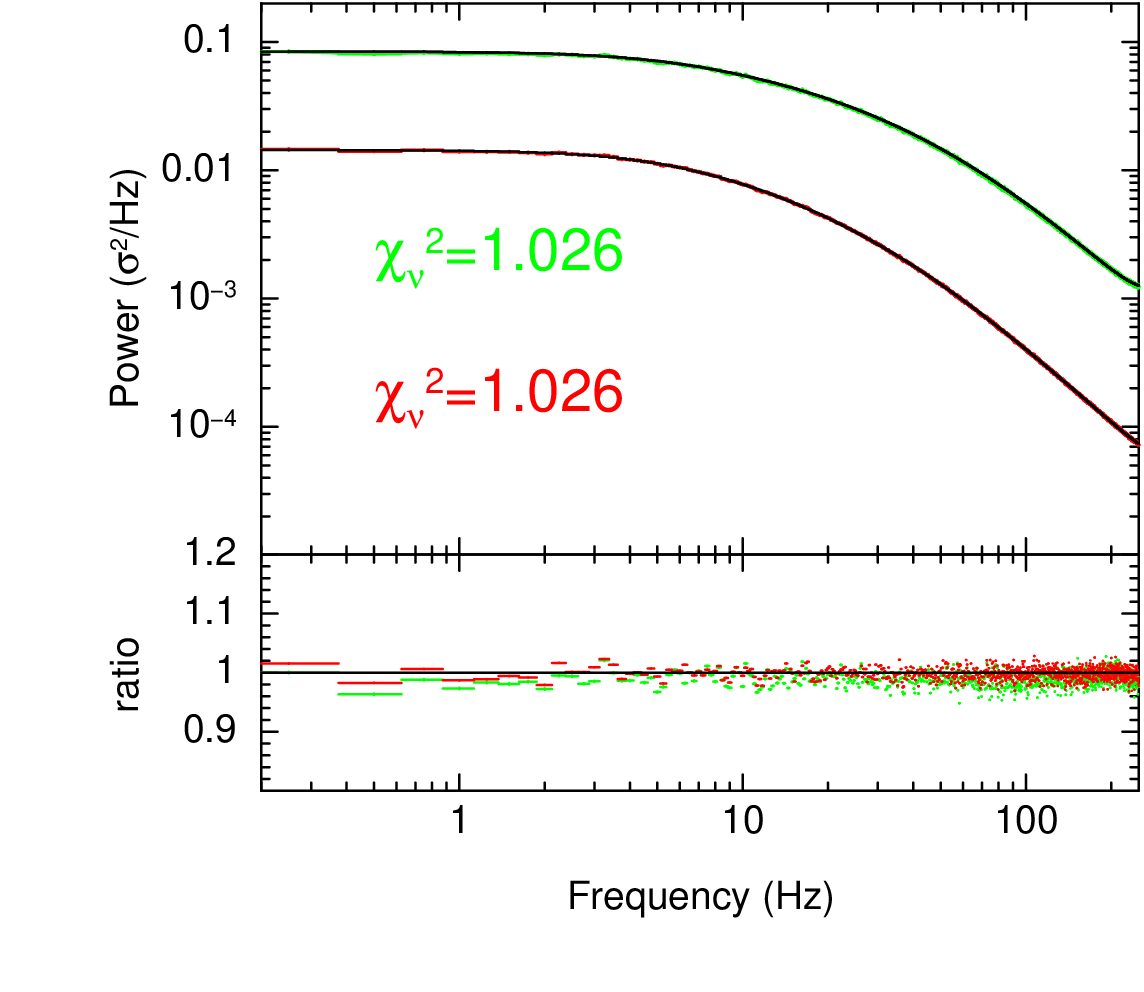}
\vspace{-10mm}
\caption{
The resulting power spectrum when 30 time series are multiplied
together (see text for details). The red and green lines
represent the simulation results when we give each time series a
variance of $\sigma = 0.5/\sqrt{30}$ and $1/\sqrt{30}$
respectively (the green points are above the red points for readers in
black and white). The black lines passing through the simulation points
are analytic calculations. The ratio plots and $\sim$unity reduced
$\chi^2$ values again confirm that the simulation and calculation agree. }
\label{fig:30rings}
\end{figure}

In Figure \ref{fig:30rings}, we consider an example with 
$\mathcal{N}=30$ functions. The input power spectra are zero-centred
Lorentzians with width, $\Delta \nu_n$, changing from  $\Delta \nu_1 =
5$ Hz to $\Delta \nu_{\mathcal{N}}=40$ Hz  and there is an equal
logarithmic spacing between the widths of consecutive Lorentzians. All
30 functions have DC components corresponding to a mean $\mu=1$ and
a standard deviation in the time domain of $\sigma =
RMS/\sqrt{\mathcal{N}}$. For the red and green points, we set
$RMS=0.5$, and 1 respectively (the $RMS=$1 points are above the
$RMS=0.5$ points). As in Figure \ref{fig:pop}, we use
$N=2^{11}$, $dt=2^{-9}$s and average the simulated power spectra over
10000 realisations. The ratio plots and $\chi^2$ statistic confirm
that the black lines calculated using equation \ref{eqn:coprod} pass
identically through the simulation points.

\section{The propagating fluctuations model}
\label{sec:mod}

\subsection{Model summary}
\label{sec:propfluc}

Here we summarise the propagating fluctuations model of
ID12 which we use in this paper with a few small alterations. We
consider a flow extending between outer and
inner radii $r_o$ and $r_i$, employing the convention that lower case
$r$ is radius expressed in units of $R_g=GM/c^2$ such that $r=R/R_g$.
We split the flow into $\mathcal{N}$ rings, each centred at $r_n$ with
an equal logarithmic spacing $dr_n$ such that $dr_n/r_n = dr/r =$
constant. We assume that fluctuations in the local mass accretion rate
are stirred up at each ring. These fluctuations are stochastic and
therefore have a random phase but they have a well defined
power spectrum given by a zero-centred Lorentzian breaking at the
local viscous frequency, i.e.:
\begin{equation}
|A_n(\nu)|^2 =
\frac{\sigma^2}{T\pi}\frac{\Delta\nu_n}{(\Delta\nu_n)^2+\nu^2},
\label{eqn:lore}
\end{equation}
where $\Delta\nu_n = 1/t_{visc}(r_n)$ and $\sigma^2$ and $T$ are the
variance and duration of the corresponding time series
respectively. We assume $\sigma$ is the same for each ring, as is the
average, $\mu$, which we set to unity. We set $\sigma/\mu =
F_{var}/\sqrt{N_{dec}}$ where $N_{dec}$ is the number of rings per
radial decade and $F_{var}$ is therefore the fractional variability
generated per decade. $F_{var}$ is thus a physical parameter of the
model and $N_{dec}$ is a parameter governing the resolution.

This amounts to a physical assumption that each radial decade in the
flow generates the same variability power. Simulations
often find that the MRI generates noise with this property
(e.g. Beckwith, Hawley, \& Krolik 2008), although we discuss an
exception in section \ref{sec:discuss}. In ID12, we used a slightly
different assumption that a constant variability per decade in viscous
frequency is generated by the MRI. This helps the model to converge
for fewer rings but the physicality of this assumption is
questionable. Here, the model is much faster to calculate and so a
high required value of $\mathcal{N}$ is no longer prohibitive, thus we
revert back to the more physically acceptable assumption adopted in
ID11, and also Arevalo \& Uttley (2006).

As in ID12, we assume a bending power law form for the time averaged
surface density, $\Sigma(r)$, which follows $\Sigma(r)\propto
r^{-\zeta}$ for $r>>r_{bw}$ and $\Sigma(r)\propto r^{\lambda}$ for
$r<<r_{bw}$. Here $r_{bw}$, the bending wave radius, is the radius
at which frame dragging torques set up plunging streams in the tilted
accretion flow simulations of Fragile et al (2007), Fragile \& Meier (2009)
and Fragile (2009). This radius is typically larger than the innermost stable
circular orbit $r_{bw} \sim 4-11$ (see Ingram, Done \& Fragile
2009). Long term mass conservation then yields an expression for the
local viscous time scale, $t_{visc}(r_n) = 2\pi
R_n^2\Sigma(r_n)/\dot{m}_0$, where $\dot{m}_0$ is the time averaged
mass accretion rate. Thus, the viscous time scale also has a bending
power law form.

The fluctuations generated in every ring will propagate inwards such
that the local mass accretion rate at $r_n$ is given by:
\begin{equation}
\dot{m}(r_n,t) = \dot{m}_0 \prod_{l=1}^{n} a_l(t-\Delta t_{ln}),
\label{eqn:mdot2}
\end{equation}
where $\Delta t_{ln}$ is the propagation time from $r_l$ to $r_n$,
given by:
\begin{equation}
\Delta t_{ln} = \frac{dr}{r} \sum_{q=l+1}^{n} t_{visc}(r_q).
\label{eqn:deltat}
\end{equation}
Thus, for example $\Delta t_{nn} = 0$ and $\Delta t_{(n-1)n} =
dr/r~t_{visc}(r_n)$. This can be alternatively written in terms of the
mass accretion rate in the $(n-1)^{th}$ ring:
\begin{equation}
\dot{m}(r_n,t) = a_n(t)\dot{m}(r_{n-1},t-\Delta t_{(n-1)n}).
\label{eqn:mdot}
\end{equation}
Thus, at the outermost ring, $r_1$, the locally generated fluctuations
will only be multiplied by the average mass accretion rate $\dot{m}_0$
whereas rings closer to the black hole will be modulated with
fluctuations generated at all outer rings.

The \textit{total} (over all energies) luminosity available to be
radiated in the $n^{th}$ ring is $\propto \dot{m}(r_n,t)$. If the
energy dependence of emission from each ring stays constant in time,
then the flux observed in some energy band can be written as:
\begin{equation}
f_h(t) = \sum_{n=1}^{\mathcal{N}} h_n \dot{m}(r_n,t),
\label{eqn:f}
\end{equation}
where $h_n$ is a set of weighting factors and we apply the convention that
$f_h(t)$ is the hard band flux (since we fit the model to the $>10$ keV
power spectrum in ID11 and ID12). The ID12 model assumes
$h_n \propto dr/r~r_n^{2-\gamma_h}b(r_n)$, with the boundary
condition $b(r_n) \propto \Sigma(r_n)$. In a truncated disc geometry, we
expect the inner, more photon starved  regions of the flow to emit a
harder spectrum than the cooler outer regions implying a steeper
emissivity for higher energy bands; i.e. $\gamma_s < \gamma_h$, where
$s$ denotes a soft band. It is this physical property that allows
the propagating fluctuations model to predict the observed lag between
hard and soft energy bands (Arevalo \& Uttley 2006; Kotov, Churazov \&
Gilfanov 2001).

\subsection{Power spectrum of the local mass accretion rate}
\label{sec:Pmdot}

We can use the result from section \ref{sec:rwalk} along with the
`time shifting' property of Fourier transforms in order to find that
the power spectrum of the mass accretion rate at $r_n$ is given by:
\begin{equation}
|\dot{M}(r_n,\nu)|^2=|A_n(\nu)|^2 \otimes |e^{i2\pi\Delta t_{(n-1)n}\nu}
\dot{M}(r_{n-1},\nu)|^2 .
\end{equation}
Note, here and for the rest of the paper we represent power spectra
(and also time series and their Fourier transforms) as continuous
rather than explicitly evaluating all power spectra at discrete
frequencies $\nu_j$. The phase shift clearly cancels here and so we
can write the power spectrum of the mass accretion rate at $r_n$ as:
\begin{equation}
|\dot{M}(r_n,\nu)|^2 = \dot{m}_0^2 \coprod_{l=1}^n |A_l(\nu)|^2.
\label{eqn:Pmdot}
\end{equation}

We compute the convolutions by transforming into the time domain
using fast Fourier transforms (FFTs), multiplying and then
transforming back with FFTs. This is by far the most computationally
efficient method to compute a convolution (Press et al 1992).

\subsection{Power spectrum for a given energy band}
\label{sec:psd}

We now calculate the power spectrum of the `hard band' flux
$f_h(t)$ (note this is a nominal choice, we have simply defined the
radial emissivity as a power law). We can transform equation
\ref{eqn:f} and take the modulus squared to show:
\begin{equation}
P(\nu) = |F_h(\nu)|^2 = \sum_{l,n=1}^{\mathcal{N}} h_l h_n
\dot{M}(r_l,\nu)^*\dot{M}(r_n,\nu),
\end{equation}
where $l$ and $n$ both take every value between $1$ and $\mathcal{N}$.
The terms with $l=n$ are easy to evaluate since they reduce to
$h_n^2|\dot{M}(r_n,\nu)|^2$. To evaluate the other terms (the cross
spectra), we must appreciate which components of the mass accretion
rate from different rings are coherent. We show in Appendix
\ref{sec:mcross} that the cross spectra are given by:
\begin{equation}
\dot{M}(r_l,\nu)^*\dot{M}(r_n,\nu) = \dot{m}_0 \Lambda_{ln}
e^{i 2\pi \Delta t_{ln}\nu} |\dot{M}(r_l,\nu)|^2,
\label{eqn:crossm}
\end{equation}
where $n > l$ and $\Lambda_{ln} \equiv \prod_{q=l+1}^{n}\mu_q$. For this
model, we always set $\mu=1$ so this product reduces to
unity. However, to preserve generality, we leave it in the resulting
equation for the power:
\begin{eqnarray}
& &P(\nu) = \sum_{n=1}^{\mathcal{N}} \bigg[ 
h_n^2 |\dot{M}(r_n,\nu)|^2  \nonumber \\
& & + 2\sum_{l=1}^{n-1} h_l h_n \Lambda_{ln} \cos(2\pi\Delta
  t_{ln}\nu) |\dot{M}(r_l,\nu)|^2 \bigg],
\label{eqn:Pf}
\end{eqnarray}
where we have used the fact that the conjugate of
$\dot{M}(r_l,\nu)^*\dot{M}(r_n,\nu)$ is
$\dot{M}(r_n,\nu)^*\dot{M}(r_l,\nu)$. Thus we have a simple sum plus
cross terms which represent interference between contributions from
each ring. The phase lag between radiation from different rings
determines whether the interference is constructive or destructive. We
see that all the cross terms will cancel if we set all $\mu_n=0$. This
makes sense physically because, in this case, the mean local mass
accretion rate inside $r_o$ is zero and thus no fluctuations can
propagate leaving radiation emitted from different rings uncorrelated.

\begin{figure}
\centering
\leavevmode  \epsfxsize=8.5cm \epsfbox{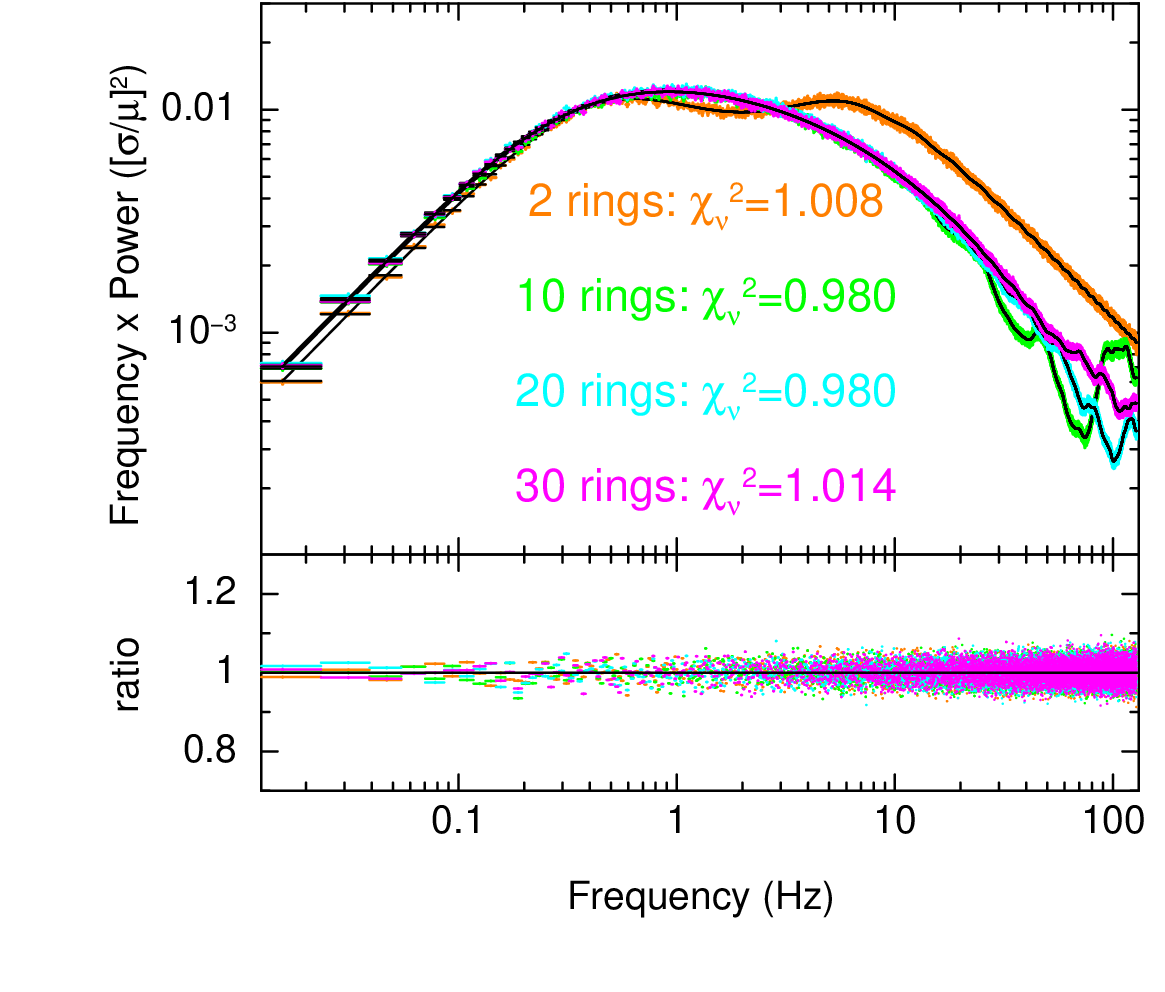}
\vspace{-10mm}
\caption{Power spectrum, normalised to show fractional variability,
  predicted from the model parameters in the text. Different colours
  represent different model resolution with $N_{dec}=$ 3, 12, 23, 35
  corresponding to 2, 10, 20 and 30 rings respectively (for readers in
  black and white, the lowest resolution model shows two distinct
  bumps and after that, higher resolutions display less oscilliatory
  behaviour at high frequency). The points
  with error represent a simulation averaged over 2000 iterations,
  whereas the black lines are calculated from equation
  \ref{eqn:Pf}. The ratio plots and reduced $\chi^2$ values confirm
  that the simulation reproduces the analytical result.}
\label{fig:Pres}
\end{figure}

In Figure \ref{fig:Pres}, we plot the power spectrum predicted by the
model using both simulation and equation \ref{eqn:Pf}. We use the best
fit parameters found for observation 3 in ID12 (except here we do not
include the QPO). We thus set $r_o=25$, $r_i=3.3$, $r_{bw}=8.7$,
$\lambda=0.9$, $\zeta=0$, $\gamma_h=5.3$ and $F_{var}=0.3$. We assume a
black hole mass and spin of $M = 10 M_\odot$ and $a_* = 0.5$ respectively
throughout this paper. The simulated power spectra are plotted with
(1 sigma) errors and the black lines show the analytically calculated power
spectra. We vary the resolution of the model showing four examples with
orange, green, cyan and magenta representing $N_{dec}=$ 3, 12, 23 and 35
respectively (for readers in black and white, see the Figure
caption). These values are chosen to give 2, 10, 20, and 30 rings 
in total. The ratio plots and $\chi^2$ values confirm that the
predicted power spectrum for a given set of model parameters is
identical whether we simulate or use equation \ref{eqn:Pf}. We also
see the effect of interference (the cosine term in equation
\ref{eqn:Pf}) on the power spectral shape for different numbers of rings.

\begin{figure}
\centering
\leavevmode  \epsfxsize=8.5cm \epsfbox{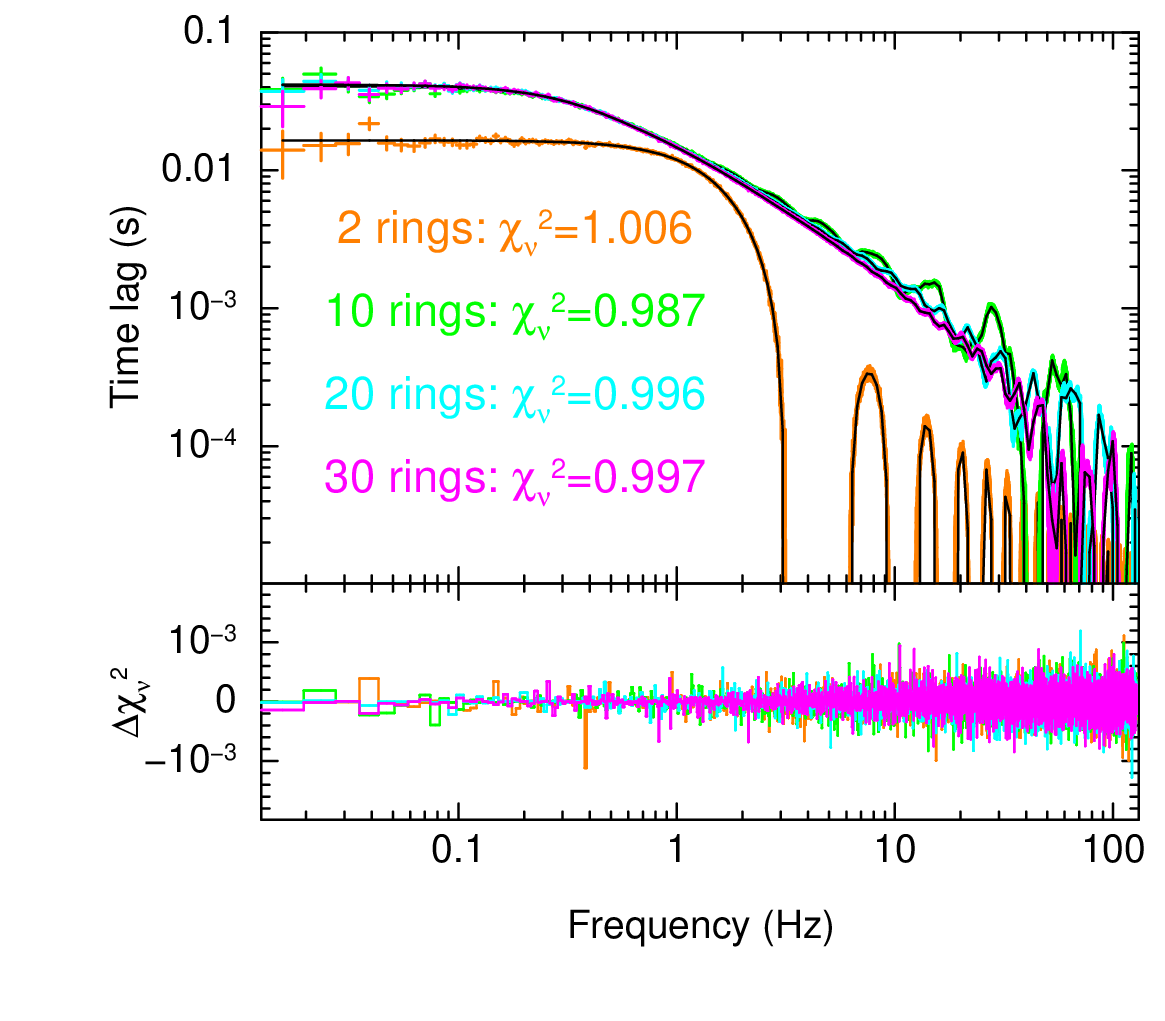}
\vspace{-10mm}
\caption{Lag spectrum predicted using the model parameters in the
  text. As in Figure \ref{fig:Pres}, different colours represent
  $N_{dec}=$ 3, 12, 23, 35, corresponding to 2, 10, 20 and 30 rings
  respectively (for readers in black and white, lower resolution
  models are associated with increasingly oscillatory
  behaviour). Again, the points  with error represent the simulation 
  and the black lines are calculated analytically, with the two showing
  excellent agreement (quantified by the $\sim$unity reduced $\chi^2$
  values). Here, we averaged the simulation over 2000 realisations.}
\label{fig:tlag}
\end{figure}

\subsection{Cross spectrum for two given energy bands}
\label{sec:cross}

In the previous section, we calculated the power spectrum of our
nominal hard band. It is clear that the power spectrum of some other
(soft) band can also be calculated from equation \ref{eqn:Pf} by using a
different emissivity (i.e. substituting $s$ for $h$). We can go
one step further and calculate the time lags between the two
bands. These can be found using the cross spectrum, $C(\nu) =
F_h(\nu)^*F_s(\nu)$ which, in contrast to the power spectrum, is
complex. The phase lag between radiation from the two bands
is given by $\tan[\Phi(\nu)] = \Im[C(\nu)] / \Re[C(\nu)]$ and the
corresponding time lag is simply $t_{lag}(\nu) = \Phi(\nu) /
(2\pi\nu)$.

It is relatively simple to adapt equation \ref{eqn:Pf} to show that
the cross spectrum is given by:
\begin{eqnarray}
& & C(\nu) = \sum_{n=1}^{\mathcal{N}} \bigg[ 
h_n s_n |\dot{M}(r_n,\nu)|^2  \\
& & + \sum_{l=1}^{n-1} (h_l s_n e^{i2\pi\Delta t_{ln}\nu}+h_ns_l
  e^{-i2\pi\Delta t_{ln}\nu} )
 \Lambda_{ln} |\dot{M}(r_l,\nu)|^2
\bigg]. \nonumber
\end{eqnarray}
This can be separated out into real and imaginary parts by splitting
the exponentials into sines and cosines to give:
\begin{eqnarray}
& & \Re[C(\nu)] = \sum_{n=1}^{\mathcal{N}} \bigg[ 
h_n s_n |\dot{M}(r_n,\nu)|^2 \label{eqn:reC} \\
& & + \sum_{l=1}^{n-1} (h_l s_n +h_ns_l) \cos(2\pi\Delta
  t_{ln}\nu)  \Lambda_{ln} |\dot{M}(r_l,\nu)|^2 \bigg]. \nonumber
\end{eqnarray}
for the real part and:
\begin{equation}
\Im[C(\nu)] = \sum_{n=1}^{\mathcal{N}} \sum_{l=1}^{n-1} (h_l s_n
-h_ns_l) \sin(2\pi\Delta t_{ln}\nu) 
 \Lambda_{ln} |\dot{M}(r_l,\nu)|^2 \label{eqn:imC}
\end{equation}
for the imaginary part.

In Figure \ref{fig:tlag}, we plot the predicted time lags using the
same model parameters as in the previous section with the additional
assumption that the soft band emissivity index is $\gamma_s =
4.5$. Again, orange, green, cyan and magenta represent 2, 10, 20 and
30 rings respectively (for readers in black and white, see the Figure
caption) and the points with error bars are from simulation
whereas the lines are calculated analytically (equations
\ref{eqn:reC} and \ref{eqn:imC}). We calculate the errors on the
simulation using the formula from Nowak et al (1999). Again, the
$\chi^2$ values confirm the simulation returns the same result as the
analytical expression. We plot contributions to $\chi^2_\nu$ (positive
means the simulation points are above the calculation) instead of
ratio because the lag spectrum passes through zero for some parameter
values. We see that at least 30 rings are required to achieve
convergence. In fact, even the 30 ring model is slightly
under-resolved with oscillatory behaviour above $\nu\approx 
10$ Hz. However, this occurs well below the Poisson noise level for
currently available observational data. This will not be the case for
the large area detector (LAD), the primary instrument of the proposed
European Space Agency mission \textit{LOFT} (the Large Observatory For
X-ray Timing: Feroci et al 2012), which will have a collecting area
$\sim$20 times that of \textit{RXTE}.

\subsection{Including the QPO}
\label{sec:qpo}

Ingram, Done \& Fragile (2009) proposed that the QPO (i.e. the type C
QPO, including all harmonics), is due to
Lense-Thirring precession of the entire flow. The model of ID12
calculates the precession frequency, $\nu_{prec}$, from the surface
density profile and the inner and outer flow radii. The power spectrum
of the QPO is then taken as a sum of Lorentzians peaking at
$\nu_{prec}$, $2\nu_{prec}$, $3\nu_{prec}$ and $\nu_{prec}/2$
representing the fundamental, $2^{nd}$, $3^{rd}$ and sub- harmonics
respectively, with the width of the fundamental $\Delta \nu_{qpo}$ left as a
free parameter. The width of the $2^{nd}$ and $3^{rd}$ harmonics are then
fixed at $2\Delta \nu_{qpo}$ and $3\Delta \nu_{qpo}$ respectively
with the width of the sub-harmonic left free (since the sub-harmonic
is often observed to have a different width; Rao et al 2010).
ID12 generate a light curve from this using the
TK95 algorithm and add this to the light curve
generated from the propagating fluctuations simulation. For
simplicity, the final light curve is normalised such that its power
spectrum is equal to the sum of the two component power spectra. This
is not particularly realistic since this normalisation implies that at
least one of the two component light curves has zero mean. However, it
provides the simplest possible way of fixing model parameters using
both the QPO frequency and the shape of the broad band noise.

For a more realistic treatment, we must consider the physical
mechanism by which the precession frequency modulates the
emission. The two most significant modulation mechanisms will be
\textit{projected area variation} as the brightest patch of the flow
moves in and out of the observer's line of sight (see Ingram \& Done
2012b) and also \textit{seed photon variation} as the flux of disc
photons incident on the flow changes as a function of precession
phase. In this paper, we consider both an additive (appropriate for
seed photon variation) and a multiplicative model (appropriate for
projected area variation). For the additive model, we use exactly the
same treatment as in ID12: the overall flux is $f_{tot}(t) =
f(t)+Q(t)$, where $Q(t)$ and $f(t)$ contain respectively the
quasi-periodic and aperiodic variability. Here, $f(t)$ has a mean of
unity, whereas $Q(t)$ has a mean of zero. Since we have assumed the
two processes to be uncorrelated, the final power spectrum will be the
sum of the powers for each process. For the multiplicative model, we
take the overall flux to be $f_{tot}(t) = f(t)Q(t)$, where $Q(t)$ now
has a mean of unity. The assumption that these two process are
uncorrelated now leads to the final power spectrum being a
convolution between the powers of each process. Note that,
in principle, we could also investigate other QPO mechanisms in this
manner. For example, we could replace one of the $\dot{m}(r_n,t)$ functions
with a quasi-periodic function $Q(t)$ to represent some oscillating
mode in a narrow region of the accretion flow.

\section{Example fits to RXTE data}
\label{sec:results}

Here we fit the model to \textit{RXTE} data from the 1998 outburst of
the transient BHB XTE J1550-564. We look at one observation from the
intermediate state of this outburst with observational ID
30188-06-01-03, referred to in ID12 as observation 3. We calculate the
white noise subtracted power spectrum of the 13.36-27 keV light curve
by averaging over 15 intervals, each containing $2^{15}$ time bins of
length $dt = 2^{-8}$s, and re-bin geometrically using a re-binning
constant of $1.05$ (van der Klis 1989). We further group bins in which
the power is averaged over less than 200 raw periodogram points. This
ensures that the statistical error on the mean periodogram has
converged to the Gaussian limit for each bin and thus the use of
$\chi^2$ as a fit statistic is appropriate.

\begin{figure}
\centering
\leavevmode  \epsfxsize=8.5cm
\epsfbox{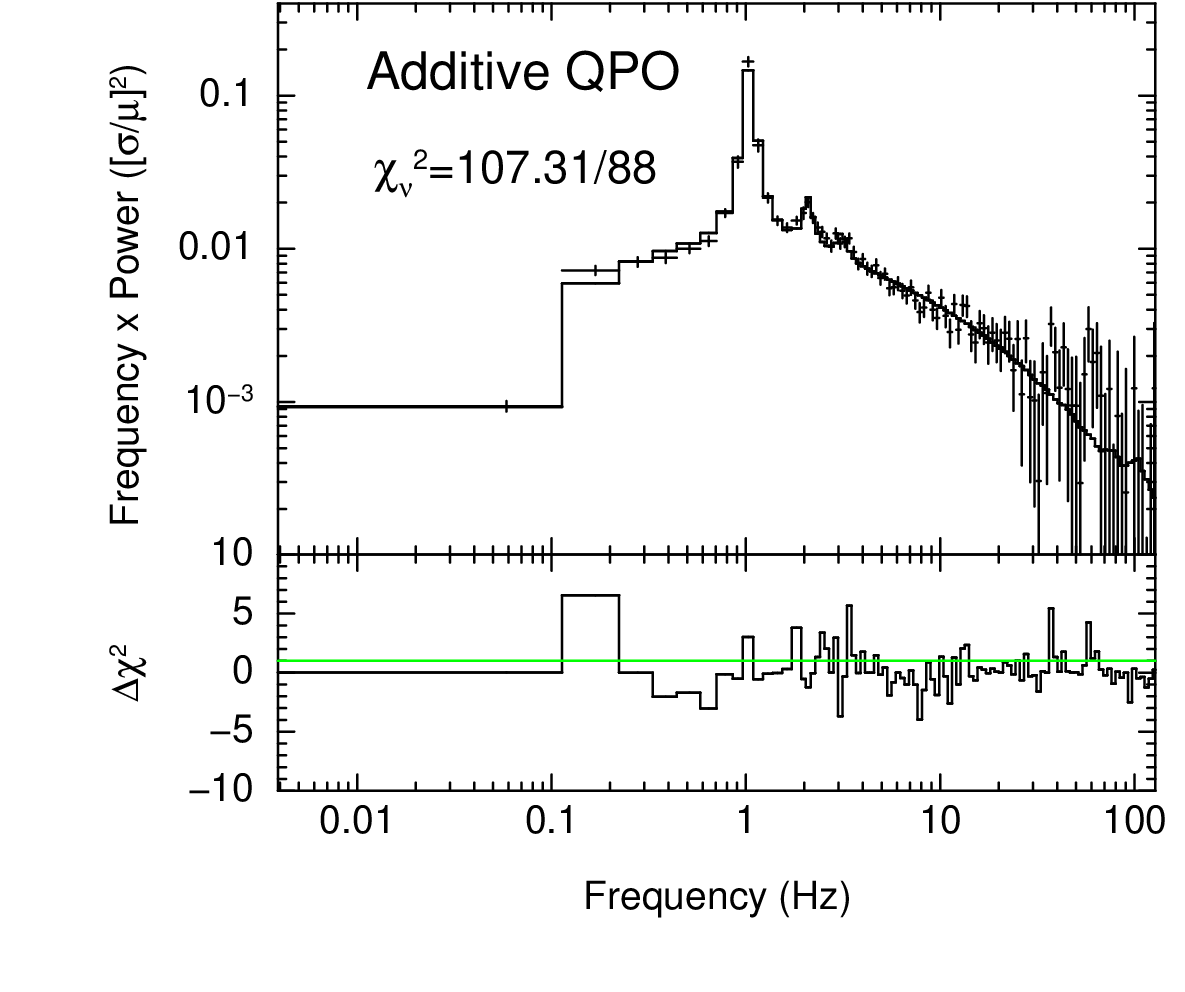}
\\
\leavevmode  \epsfxsize=8.5cm
\epsfbox{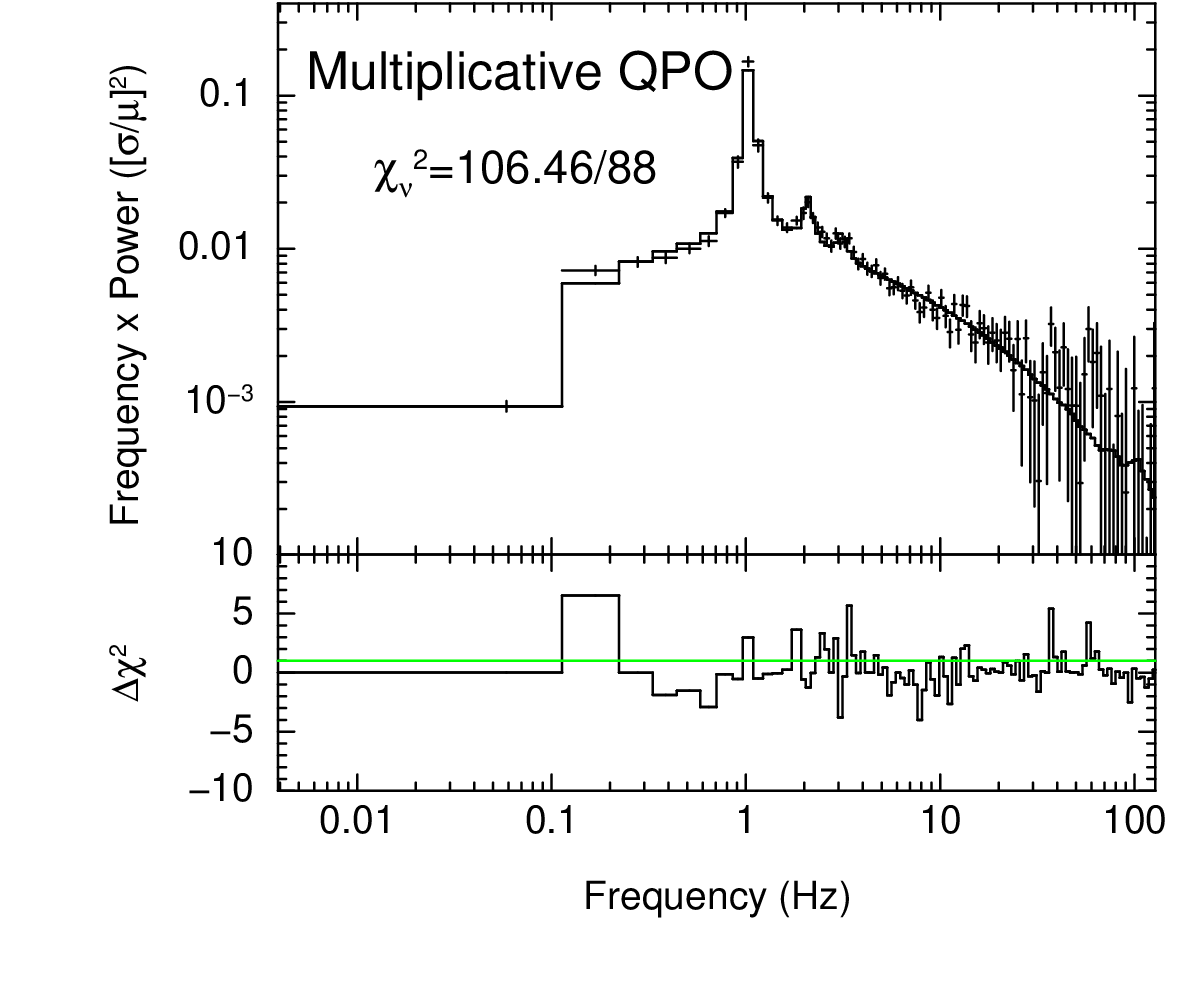}
\vspace{-10mm}
\caption{
White noise subtracted 13.36-27 keV power spectrum of observational ID
30188-06-01-03 with best fit additive (top) and multiplicative
(bottom) models (see Table \ref{tab:results}).}
\label{fig:results}
\end{figure}

We first fit the additive model, which treats the QPO in the same way
as ID12, setting $N_{dec}=30$. We use \textsc{xspec} version 12
(Arnaud, Borkowski, \& Harrington 1996) to find the best least-squares
fit, freezing all the parameters that were held constant in
ID12. Figure \ref{fig:results} (top) shows the result plotted in
frequency $\times$ power with the contributions to $\chi^2$ also
plotted underneath. The best fit parameters, shown in Table
\ref{tab:results}, give a minimum $\chi^2_\nu=107.31/88=1.22$. We also
fit the multiplicative model for which the QPO power spectrum is
convolved with that of the broad band noise instead of added. We find
the fit plotted in Figure \ref{fig:results} (bottom) is slightly
better ($\chi^2_\nu=106.46/88=1.21$), although the difference is not
significant. The main difference between the two sets of parameters is
$F_{var}$ which is smaller for the multiplicative model because here
the final power spectrum is a sum of the broad band noise, the QPO and
the convolution between the AC components of these two components (see
equation \ref{eqn:dc}) which enhances the total variability predicted
for a given value of $F_{var}$.

We simulate both versions of the model with 1000 realisations to
confirm convergence to the analytic calculation. For the additive and
multiplicative models we find agreement between simulated and
calculated power with $\chi_\nu^2=1.0164$ and $\chi_\nu^2=1.014$
respectively (both for $2^{15}$ degrees of freedom). We also simulate
using only 15 realisations to directly compare the simulation to
data. When we use the binning scheme described above, we find that the
simulation does converge to the analytic calculation with reduced
$\chi^2$ values of 1.004 and 1.014 respectively for the additive and
multiplicative versions of the model (both for 97 degrees of freedom).
Thus, if our model did perfectly describe the observed data, it
would indeed return a reduced $\chi^2$ of unity. Table
\ref{tab:results} also shows the result of re-fitting the ID12 model
using the analytic formulation. The new best fit parameters are close
to those obtained in ID12 using the simulation method (with the
fitting algorithm taking minutes to run for the analytical calculation
as opposed to weeks for the old simulation method) and give a reduced
$\chi^2$ value of $\chi^2_\nu=120.96/88=1.37$.

It is possible to see from Figure \ref{fig:results} that both models
predict the slope between $\sim0.1$ and $1$ Hz to be too steep (this
can also be seen in the contributions to $\chi^2$). In the hard state of
most BHBs when there is no (or only a weak) QPO present, it is
possible to see by eye that the broad band noise is best modelled by
two (or perhaps more) discrete bumps rather than the `flat top' noise
predicted by this model. It is possible that the broad band noise in
this intermediate state power spectrum is also best modelled with two
discrete bumps, resulting in the small discrepancy between our model
and the data. We will discuss the potential physical implications of this in
the following section.

Although the  power spectra for the additive
and multiplicative models are only subtly different, the predicted
bi-coherence (Maccarone \& Coppi 2002) will be extremely
different. This is a measure of correlation between different Fourier
frequencies and reveals that the QPO correlates strongly
with the broad band noise in GRS 1915+105 (Maccarone et al 2011) and
also for the observation of XTE J1550-564 considered here (Tom
Maccarone; private communication). This strong correlation can not be
reproduced by the additive model but potentially could by the
multiplicative model, strongly favouting the latter.

\begin{table}
\begin{tabular}{l|l|l|l|c}
 \hline
  Parameter & Additive QPO & Multiplicative QPO  & ID12 \\
 \hline
 \hline
$\Sigma_0$ ($\dot{M}_0/[cR_g]$)   & $27.34\pm 3.37$   & $  28.09 \pm 2.86$ & $23.67\pm 4.45$\\
$r_{bw}$ ($R_g$)                             & $10.86\pm 1.60$   & $11.21 \pm 1.80$ & $6.80\pm1.10$\\
$\kappa$                & $\equiv 3$     & $\equiv 3$ &$\equiv 3$ \\
$\lambda$              & $\equiv 0.9$   & $\equiv 0.9$ &$\equiv 0.9$\\
$\zeta$                    & $\equiv 0$  & $\equiv 0$ &$\equiv 0$\\
$F_{var}$                   & $0.276 \pm 0.01$  & $0.138\pm 0.008$ & $0.372\pm0.018$\\
$r_o$ ($R_g$)            & $24.422 \pm 0.48$   & $24.32\pm 0.10$ & $26.03\pm0.49$\\
$r_i$ ($R_g$)             & $\equiv 3.3$   & $\equiv 3.3$ &$\equiv 3.3$ \\
$\Delta\nu_{QPO}$ ($10^{-2}$Hz)   &  $7.05\pm 0.46$   & $7.03 \pm 0.46$ &$6.12\pm0.47$\\
$\sigma_{qpo}$  ($10^{-2}$)         & $17.16 \pm 0.46$   & $17.13\pm 0.47$ & $17.18\pm0.47$ \\
$\sigma_{2qpo}$ ($10^{-2}$)         & $4.94 \pm 0.27$   &  $4.93\pm  0.27$  &$4.80\pm0.25$\\
$\sigma_{3qpo}$ ($10^{-2}$)         & $2.94 \pm 0.26$   & $ 2.93\pm  0.28$&$2.78\pm0.26$\\
$\gamma$             & $5.39 \pm 0.27 $  & $5.56\pm  0.37$ & $4.97\pm0.53$ \\
$M$     ($M_\odot$)                  & $\equiv 10 $   & $\equiv 10 $ &$\equiv 10 $\\
$a$                        & $\equiv 0.5$              & $\equiv 0.5$&$\equiv 0.5$ \\
$\chi^2_\nu$                        & $107.31/88$              & $106.46/88$& $120.96/88$ \\
\hline
\end{tabular}
\caption{
Best fit parameters with associated ($1\sigma$) error estimates for
the additive and multiplicative models (see Figure \ref{fig:results})
alongside those from ID12.}
\label{tab:results}
\end{table}     

\section{Discussion \& Conclusions}
\label{sec:discuss}

The model of ID11 and ID12 combines propagating mass accretion rate
fluctuations with Lense-Thirring precession of the inner flow to fit a
physical model to a BHB power spectrum for the first time. In these
papers, and previous studies of the propagating fluctuations model
(Arevalo \& Uttley 2006), a Monte Carlo simulation is used to generate
stochastic light curves and the power spectrum is \textit{estimated} by
averaging over many realisations. This process is heavily
computationally intensive and inevitably leads to an inexact model
with an associated error estimate. Here, we calculate the
\textit{same} model \textit{exactly} by deriving an analytic
formula. We find that, in this context, the power of mass accretion
rate fluctuations from a given region of the flow at each Fourier
frequency is a random walk on the complex plane and a standard
statistical result gives an exact expression. We then derive an exact
expression for the power spectrum of \textit{any} linear combination
of these mass accretion rate functions. We can thus calculate the ID12
model exactly. We fit our model to an observation previously
considered in ID12. We can now, in contrast to ID12, fully explore
$\chi^2$ space and run error calculations. We obtain a fairly good fit
with residuals which hint at changes we must make to our physical
assumptions going forward. We also find that the more physical
assumption of a multiplicative QPO gives a marginally better fit than
the previously assumed additive QPO.

The model here still ignores intrinsic disc variability. Since this
has been observationally confirmed in the hard state (Wilkinson \&
Uttley 2009), it will be more appropriate to consider noise to be
generated in both the disc and flow with a discontinuity in viscous
time scale occurring at the truncation radius (since this will be
shorter in the flow than in the disc). This discontinuity in
viscous time scale will allow the model to reproduce the double hump
power spectra often observed in the hard state and may even improve
the fit for intermediate state power spectra such as the observation
considered here. The model also assumes that the MRI generates equal
variability power per decade in radius. In ID12, we speculated that
frame dragging torques could give rise to enhanced variability in the
bending wave region. We showed that, for this effect to give rise to a
bumpy power spectrum, the variability in the bending wave region needs
to be a factor of $\sim 10$ greater than elsewhere (Figure 6
therein). Henisey, Blaes \& Fragile (2012) have since found that the
tilted accretion flow GRMHD simulations of Fragile et al (2007) do
indeed show enhanced variability in the bending wave region (see
Figure 7 therein) a factor $\sim 10$ greater than elsewhere! Thus 
we may need to consider this going forward.

There are a number of other obvious improvements we can make to the
model. We stress that our result here is \textit{very}
powerful since it can still be used with far more sophisticated versions
of the model than the one considered here. First of all, the ID12
model effectively assumes an overly simplified form for the power
spectrum of the Green's function (the Green's power) of the flow. The true
Greens function for a Keplerian flow can be found from the diffusion
equation (Lynden-Bell \& Pringle 1974; Pringle 1981; Frank, King \&
Raine 2002; King et al 2004) and, providing the system is linear, can
be convolved with any intrinsic fluctuations generated by, say, the
MRI to give the resulting fluctuations in mass accretion rate. In
Fourier space, this is a multiplication and thus if the input
fluctuations are white noise, the power spectrum of the mass accretion
rate simply becomes the Green's power. In general, the Green's power
is a function $G(r_l,r_n,\nu)$, where $r_l$ and $r_n$ are respectively
the radii at which the fluctuation was generated and reacted to. In the limit
$r_l>>r_n$, the Green's power becomes a zero-centred Lorentzian 
with width $1/t_{visc}(r_n)$ (equation \ref{eqn:lore}). Since the
analytic formulae we derive here are appropriate for \textit{any}
Green's function, we will apply it in future to a more general Green's
function derived directly from the diffusion equation.

Also, we have thus far assumed that the Compton up-scattering process
which creates the power law emission is effectively instantaneous. In
reality, this process will also contribute a Green's function with a
power spectrum that looks like a low-pass filter with a break at
fairly high frequency. Ishibashi \& Courvoisier (2012) speculate that
the high frequency break in the observed power spectrum could be
associated with the Compton cooling time scale in the flow. If this
timescale dictates the break frequency in the Compton Green's power,
it will indeed govern the observed high frequency break in the power
spectrum. Since this time scale naturally predicts McHardy et al's
(2006) empirical relation with mass and accretion rate, it is a very
attractive suggestion. We will investigate this in a future paper.
Also, we assume that the flux in a given energy band is a linear
combination of the mass accretion rate at each radius. In reality,
this will not be completely true since the spectral \textit{shape} of
emission from each ring will vary with mass accretion
rate. Gierlinski \& Zdziarski (2005) studied the effect of varying
certain parameters of a Comptonisation model. We plan to include a
similar analysis in a future version of our model.

In conclusion, we have shown that the power spectral model of ID12 can
be calculated analytically. We have also shown simple ways of
calculating the predicted power spectrum for different energy bands
and even the time lag between energy bands. Going forward, we can thus
fit simultaneously to all of these observational properties, utilising
the wealth of information locked in the spectral timing properties
of the X-ray data.

\section{Acknowledgements}

We acknowledge the anonymous referee for very useful comments
facilitating the clarity and swift completion of this work. This
research has made use of data obtained through the High Energy
Astrophysics Science Archive Research Center Online Service, provided
by the NASA/Goddard Space Flight Center.


\appendix

\section{Complex conjugate symmetry}
\label{sec:evilterms}

The derivation of equation \ref{eqn:proof} relies on the assumption
that the two time series being multiplied together, $a_k$ and $b_k$,
are completely uncorrelated. That this is indeed the case is not as
trivial as it first seems since the fact that $a_k$ and $b_k$ are real
introduces a correlation between $A_j$ and $A_{-j}$ (complex
conjugate symmetry). To illustrate this point, consider $a_k$ and
$b_k$ with only $N=4$ terms. Using equation \ref{eqn:cycconv}, we
know that the $j=2$ entry of the DFT of $x_k=a_k b_k$ obeys:
\begin{eqnarray}
\sqrt{4dt} X_2 &=& |A_{-1}||B_{-1}|e^{i(\alpha_{-1} + \beta_{-1} )}
                       + |A_{2}||B_{0}|e^{i(\alpha_{2} + \beta_{0} )}
                       \nonumber \\
                & & + |A_{1}||B_{1}|e^{i(\alpha_{1} + \beta_{1} )}
                       + |A_{0}||B_{2}|e^{i(\alpha_{0} + \beta_{2} )}.
\label{eqn:Xsum}
\end{eqnarray}
Here $\alpha_j$ and $\beta_j$ represent the phase of $A_j$ and $B_j$
respectively. Since $\alpha_{-j}=-\alpha_{j}$ and
$\beta_{-j}=-\beta_{j}$, we see that the $A_{-1}B_{-1}$ and $A_{1}B_{1}$
terms are correlated with each other such that:
\begin{equation}
A_{-1}B_{-1} + A_{1}B_{1} = 2|A_{1}||B_{1}| \cos(\alpha_1+\beta_1).
\end{equation}
Thus, these two terms actually behave in the sum (equation
\ref{eqn:Xsum}) as one larger term. When we then calculate $|X_2|^2$,
the sum will contain one term of size $4|A_{1}|^2|B_{1}|^2
\cos^2(\alpha_1+\beta_1)$ rather than two terms with combined size
$2|A_{1}|^2|B_{1}|^2$. However, when we calculate $<|X_2|^2>$, we find
$<\cos^2(\alpha_1+\beta_1)>=1/2$ and therefore this one large term
contributes $2|A_{1}|^2|B_{1}|^2$; the same as the total contribution
of two uncorrelated terms. For general $N$,  $A_{-j} B_{-j}$
correlates with $A_{j} B_{j}$ and $A_{-j} B_{j}$ correlates with
$A_{j} B_{-j}$ in the same manner. All of these correlations reduce in
the way demonstrated here, and thus equation \ref{eqn:cycconv} can
indeed be treated as a random walk on the complex plane.


\section{Cross spectrum calculation}
\label{sec:mcross}

The mass accretion rate in the $l^{th}$ ring is:
\begin{equation}
\dot{m}(r_l,t) = \prod_{q=1}^{l} (\tilde{a}_q (t-\Delta t_{ql}) + \mu_q),
\end{equation}
where a tilde denotes zero mean. We can write this as:
\begin{equation}
\dot{m}(r_l,t) = \sum_\theta \prod_{q=1}^{l} \theta_{ql}, 
\end{equation}
where $\theta_{ql}$ can either be $\mu_q$ or $\tilde{a}_q(t-\Delta t_{ql})$ and the sum
is over every \textit{combination} (i.e. not permutation) of
$\theta$. For example, for the third ring we have:
\begin{eqnarray}
\dot{m}(r_3,t) &=& \tilde{a}_1 (t-\Delta t_{13}) \tilde{a}_2 (t-\Delta
t_{23}) \tilde{a}_3 (t) \nonumber \\ 
&+& \mu_1 \tilde{a}_2(t-\Delta t_{23}) \tilde{a}_3 (t)
+ \mu_2 \tilde{a}_1(t-\Delta t_{13}) \tilde{a}_3 (t) \nonumber \\
 &+& \mu_3 \tilde{a}_1(t-\Delta t_{13}) \tilde{a}_2 (t-\Delta t_{23})
+ \mu_1\mu_2 \tilde{a}_3 (t)  \nonumber \\
&+& \mu_1\mu_3 \tilde{a}_2 (t-\Delta
t_{23}) + \mu_2\mu_3 \tilde{a}_1 (t-\Delta t_{13})  \nonumber \\
&+& \mu_1\mu_2\mu_3 
\end{eqnarray}
and so each term is a product of $l=3$ terms and the sum is over every
different combination that triplet can take. We can use this to know
which terms correlate between the $n^{th}$ and $l^{th}$ ring because
the mass accretion rate in the $n^{th}$ ring is:
\begin{eqnarray}
\dot{m}(r_n,t) &=& \sum_\theta \prod_{q=l+1}^{n} \theta_{qn}
\prod_{q=1}^{l} \theta_{qn} \nonumber \\
&=& \prod_{q=l+1}^{n} \mu_q \sum_\theta \prod_{q=1}^{l} \theta_{qn} +
~uncorrelated~terms. \nonumber
\end{eqnarray}
We know that $\theta_{qn}$ represents either $\mu_q$ or
$\tilde{a}_q(t-\Delta t_{qn})$. In the former case, it is easy to see
that $\theta_{qn}=\theta_{ql}$ (i.e. $\mu_q = \mu_q$). The latter case
is a little more complicated but we can use the definition of $\Delta
t_{ln}$ (equation \ref{eqn:deltat}) in order to show that
$\tilde{a}_q(t-\Delta t_{qn})= \tilde{a}_q(t-\Delta t_{ql}-\Delta
t_{ln})$ and thus:
\begin{equation}
\dot{m}(r_n,t) = \Lambda_{ln} \dot{m}(r_l,t-\Delta t_{ln}) +~uncorrelated~terms.
\end{equation}
From here it is simple to show that equation \ref{eqn:crossm} is true.

\label{lastpage}

\end{document}